\documentclass[11pt]{article}
\usepackage[a4paper,margin=1in]{geometry}
\usepackage[T1]{fontenc}
\usepackage[utf8]{inputenc}
\usepackage{lmodern}
\usepackage{microtype}
\usepackage{amsmath,amssymb}
\usepackage{booktabs,tabularx,array}
\usepackage{enumitem}
\usepackage{xcolor}
\usepackage{xurl}
\usepackage[numbers,sort&compress]{natbib}
\setcitestyle{square,comma,numbers,sort&compress}
\usepackage[colorlinks=true,linkcolor=blue!60!black,citecolor=blue!60!black,urlcolor=blue!60!black]{hyperref}
\usepackage{tikz}
\usetikzlibrary{arrows.meta,positioning,fit,calc}
\usepackage{caption}
\hypersetup{
  pdftitle={Routed Closure: Rethinking Value Capture in Decentralized Ecosystems},
  pdfauthor={Xubin Luo},
  pdfkeywords={decentralized ecosystems; tokenomics; routed closure; pooled capture; route-admissible value; EVRC; value capture}
}
\setlist{nosep,leftmargin=2em}
\newcolumntype{Y}{>{\raggedright\arraybackslash}X}
\newcommand{\EVRC}{\textsc{EVRC}}
\DeclareRobustCommand{\RAV}{\ifmmode\operatorname{RAV}\else\textsc{RAV}\fi}
\DeclareRobustCommand{\RCR}{\ifmmode\operatorname{RCR}\else\textsc{RCR}\fi}
\DeclareRobustCommand{\NCD}{\ifmmode\operatorname{NCD}\else\textsc{NCD}\fi}
\DeclareRobustCommand{\W}{\ifmmode W\else\textit{W}\fi}

\title{Routed Closure:\\
Rethinking Value Capture in Decentralized Ecosystems}
\author{Xubin Luo\\
School of Computer and Artificial Intelligence\\
Southwestern University of Finance and Economics\\
\texttt{xubinluo@gmail.com}}
\date{May 25, 2026\\
\small Working paper: framework and bounded comparative diagnostics}

\begin{document}
\maketitle

\begin{abstract}
Value-capture analysis asks where value is appropriated. In centralized firms and platforms, this question often relies on a reservoir assumption: captured value enters a governable pool and can be reallocated through payroll, contracts, budgets, subsidies, pricing, or managerial discretion. We call this centralized baseline \emph{Pooled Capture}. Decentralized ecosystems weaken that assumption. Captured value can land in a front end, issuer, treasury, token holder, burn mechanism, protocol contract, or secondary market without becoming payment to the actors who maintain the system. This paper proposes \emph{Routed Closure} as a post-capture diagnostic: captured value supports decentralized sustainability only when it passes a route-admissibility test to a specified critical incentive recipient and then a coverage-adequacy test against the relevant reward denominator. We introduce Route-Admissible Value (\RAV) as the accepted numerator for this test and use the External Value Routing Closure (\EVRC) protocol to code value landing, route enforceability, recipient specificity, evidence grade, and claim gates. A theory-driven contrast set--YouTube, Steem/Steemit, Bitcoin, Ethereum, Aave, Filecoin, USDC, and XRP--illustrates why users, fees, token burns, issuer income, or market capitalization should not be conflated with externally funded critical incentives. The paper is a framework and protocol contribution with bounded comparative diagnostics. It does not report final cross-case metric rankings, does not claim that Steemit had no revenue, and does not infer long-run Bitcoin security failure from bounded fee-share evidence.
\end{abstract}


\section{Introduction}

Decentralized ecosystems can have users, transactions, fees, treasuries, token burns, issuer balance sheets, or high market capitalization while still failing to route external use value to the people whose work keeps the system functional. A content network must pay authors and curators. A proof-of-work chain must pay miners. A proof-of-stake chain must pay validators. A lending protocol must compensate suppliers and risk-bearing layers. A storage network must pay providers. In this paper, these critical participants are denoted by \W.

The central question is:

\begin{quote}
\textbf{Does external, non-investment payment reach the critical incentive layer through a verifiable route, and does it cover the ongoing rewards required to keep the system working?}
\end{quote}

This is not the same as asking whether a project has revenue, users, or a token price. It is also not a claim that every decentralized system must mimic a centralized firm's accounting. Rather, decentralized systems lack a default unified revenue pool. If value lands at front-ends, operating companies, treasuries, burn mechanisms, issuers, or secondary markets, analysts must show how it reaches \W{} before treating it as reward coverage.

This paper calls the required condition \emph{Routed Closure}. The problem is not that value-capture evidence is useless. Value capture estimates potential value. Route admissibility decides whether that value counts as funding for a specified \W. Coverage adequacy then decides whether the admissible routed value is enough. Captured value can therefore be real and still fail the routed-closure test.

Steem/Steemit motivates the problem without exhausting it. Steemit was among the first blockchain social platforms to promise cryptocurrency rewards for contribution, while later company-level updates discussed STEEM sales, cost reductions, and advertising-related activity. The decisive economic question is not whether the internal reward curve was fair or whether the company had activity. It is whether operating revenue funded the protocol-level content reward loop, or merely coexisted with a separate issuance-funded loop.

This paper makes three contributions. First, it develops Routed Closure as a framework for moving from pooled capture assumptions to route-verified reward funding. Second, it introduces Route-Admissible Value (\RAV) and operationalizes the framework through External Value Routing Closure (\EVRC), a coding protocol that fixes the analysis unit, \W, payment motive, value landing, routing strength, evidence grade, and claim gate before any closure claim is made. Third, it applies the protocol to a theory-driven contrast set: YouTube as a centralized revenue-pool baseline; Steem/Steemit as an app--protocol fracture; Bitcoin and Ethereum as security-budget cases; Aave and Filecoin as service-payment contrasts; and USDC and XRP as issuer-closure and burn-mismatch boundaries.

The strongest claim is mechanistic, not prophetic:

\begin{quote}
\emph{Centralized value capture presumes pooling; decentralized reward closure requires routing. Application-layer value, user activity, protocol fees, token burns, and token rewards do not imply sustainability unless external use-oriented value is route-admissible for the incentive loop that pays the system's critical participants.}
\end{quote}

The contribution is a diagnostic lens, not a prediction model. We do not claim that Steemit had no revenue. We do not prove universal historical route-null from the currently captured sources. We do not forecast token prices or collapse dates. We do claim that conflating front-end or company revenue with protocol reward closure is a fundamental analytical error.

\begin{figure}[htbp]
\centering
\begin{tikzpicture}[
  every node/.style={font=\footnotesize},
  box/.style={draw, rounded corners=2pt, align=center, text width=2.34cm, minimum height=0.92cm, fill=blue!4},
  gate/.style={draw, rounded corners=2pt, align=center, text width=2.34cm, minimum height=0.88cm, fill=green!7},
  warn/.style={draw, rounded corners=2pt, align=center, text width=1.92cm, minimum height=0.76cm, fill=red!6},
  arr/.style={-{Latex[length=2.2mm]}, thick},
  darr/.style={-{Latex[length=2.2mm]}, thick, dashed, red!70!black}
]
\node[box] (pay) at (0,0) {External use payment\\{\scriptsize consumption, transaction, service fee}};
\node[box] (landing) at (3.05,0) {Value landing\\{\scriptsize app, protocol, treasury, burn, market}};
\node[gate] (route) at (6.10,0) {Verifiable route\\{\scriptsize contract, governance, protocol rule}};
\node[box] (w) at (9.15,0) {Critical workers \W\\{\scriptsize authors, miners, validators, suppliers}};
\node[gate] (coverage) at (12.20,0) {Reward coverage\\{\scriptsize enough to sustain \W?}};
\draw[arr] (pay) -- (landing);
\draw[arr] (landing) -- (route);
\draw[arr] (route) -- (w);
\draw[arr] (w) -- (coverage);

\node[warn] (b1) at (0,-2.25) {B1\\pseudo-consumption};
\node[warn] (b2) at (3.05,-2.25) {B2\\app--protocol fracture};
\node[warn] (b3) at (6.10,-2.25) {B3\\burn/capture mismatch};
\node[warn] (b4) at (9.15,-2.25) {B4\\issuance/\\market\\dependence};
\draw[darr] (b1.north) -- (pay.south);
\draw[darr] (b2.north) -- (landing.south);
\draw[darr] (b3.north) -- (route.south);
\draw[darr] (b4.north) -- (w.south);
\end{tikzpicture}
\caption{The Routed Closure diagnostic chain. Value-capture analysis may identify value landing or appropriation; \EVRC{} continues to the route-admissibility and coverage-adequacy tests for \W.}
\label{fig:chain}
\end{figure}
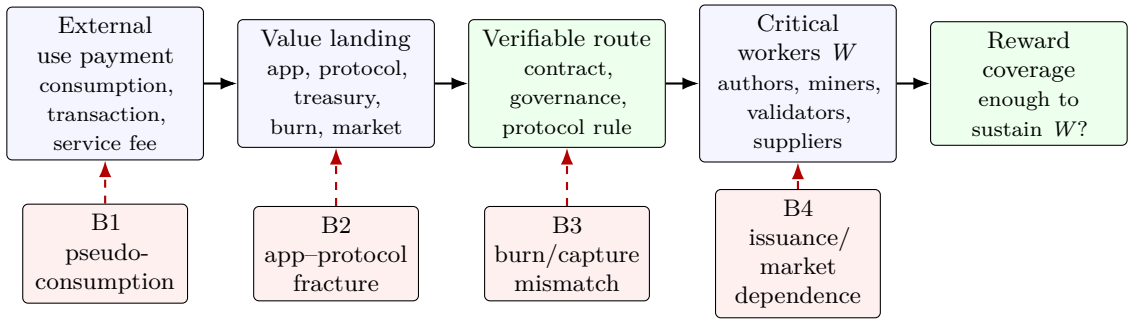

\section{From Pooled Capture to Routed Closure}

Centralized platforms maintain a unified revenue object--advertising, subscriptions, commissions, take rates--and then redistribute to creators, merchants, infrastructure providers, or operators by contract, policy, or managerial discretion. The split may be unfair, but the revenue pool is identifiable. This paper calls that background condition the \emph{reservoir assumption}: captured value is presumed to enter a governable pool from which continuing production, maintenance, subsidies, and rewards can be funded. Traditional value-capture theory often rests on this convergence structure: even when production is distributed, value capture remains organized around a focal firm, platform owner, asset holder, or other identifiable profit sink.

Decentralized ecosystems weaken that assumption. The issue is not that their value flows are always more complex. Some routes are simple and protocol-enforced. The issue is that pooling and fungible reallocation cannot be presumed from value capture alone. Value may land in front-end operators, protocol treasuries, DAO accounts, token burns, staking dilution, L2 sequencers, issuer balance sheets, or secondary-market prices while the critical incentive recipient sits at another node. These nodes are not automatically interchangeable.

Steemit illustrates the fracture. Company-level advertising or finance activity can exist at the application layer while the protocol-level reward pool distributes newly issued tokens according to blockchain rules. Treating ``the platform has a business'' as equivalent to ``authors were paid by external cash'' skips the decisive routing step.

The analytical task is therefore two-stage. First, captured value must pass a route-admissibility test: can it be counted as funding for the specified critical incentive recipient? Second, once a route is verified, the analyst can ask the coverage question: is the routed value sufficient relative to the relevant reward denominator? Value capture informs the second question only after the first question is answered.

Table~\ref{tab:symptoms} distinguishes downstream mechanism questions from the prior payer question.

\begin{table}[htbp]
\centering
\caption{Internal mechanism questions versus the prior funding-route question}
\label{tab:symptoms}
\small
\begin{tabularx}{\textwidth}{YY}
\toprule
Downstream symptom-level question & Prior funding-route question \\
\midrule
How should bots be prevented from gaming votes? & Who pays for rewards posted to \W? \\
How should whale influence be limited? & Does advertiser, subscriber, or user cash reach the reward pool? \\
How should the reward curve be tuned? & Is reward purchasing power externally funded or token-market dependent? \\
How should witnesses or validators be selected? & Are operators paid by routed use value or by issuance and market absorption? \\
\bottomrule
\end{tabularx}
\end{table}

When reward purchasing power is not backed by route-admissible external cash flow, it tends to track token issuance, market narratives, and new market absorption. If that support weakens, nominal rewards can still be paid while real purchasing power for creators, validators, miners, or providers falls. This is a structural fragility channel; it need not be labelled polemically to matter.

\section{The Routed Closure Framework}

Routed Closure builds on adjacent work on token valuation, business-model value
creation and capture, and security budgets, but narrows the unit of analysis
before making any closure claim \cite{cong2021,amit2001,teece2010,jacobides2018,weking2020,budish2018,easley2019,huberman2021}. It does not ask only whether value exists somewhere in an ecosystem. It asks whether value is route-admissible for \W{} and whether that admissible value covers the rewards required by \W. \EVRC{} names the coding protocol that operationalizes this test.

\subsection{Three Separations}

Routed Closure starts with three separations.

\begin{enumerate}
  \item \textbf{Value generation is not value landing.} Posts, transactions, storage demand, or borrowing demand may generate value, but the payment may land at an app, firm, issuer, burn mechanism, treasury, or secondary market.
  \item \textbf{Value landing is not critical incentive payment.} Company revenue, issuer reserve income, or token burn can capture value without paying the workers who maintain the system.
  \item \textbf{Critical incentive payment is not project success.} \EVRC{} diagnoses whether income reaches and covers \W. It does not predict price, governance quality, or survival.
\end{enumerate}

\subsection{Two-Stage Test}

\textbf{Route admissibility} asks whether external non-investment payment can be counted as funding for a specified \W{} through traceable and enforceable rules. It is a numerator-admission test: captured value that fails this gate remains potential value, not reward funding.

\textbf{Coverage adequacy} asks whether the route-admissible value is large, stable, and frequent enough to cover ongoing rewards, costs, and risks for \W.

The distinction is essential. Bitcoin has a protocol-enforced fee route to miners, but fee coverage remains an empirical question. Ethereum has a fee mechanism, but base-fee burn, priority fees, MEV, issuance, and penalties must be separated. Aave has service-payment routes to suppliers and protocol revenue, but that does not automatically close token-holder value or tail-risk coverage.

\subsection{Value Capture, Value Flow, and Protocolized Routing}

Routed Closure is not a replacement for value-capture theory. Traditional value-capture and business-model work asks how a firm, platform, or ecosystem actor creates, controls, appropriates, and allocates value \cite{amit2001,teece2010,jacobides2018,weking2020}. These theories usually carry an implicit convergence structure: system boundaries are relatively clear, a focal organization is relatively identifiable, and value flows can be rearranged through pricing, commissions, contracts, access rules, subsidies, or internal budgets. Even when value is created by users, developers, or complementors, capture commonly remains organized around an identifiable profit sink.

Decentralized systems change the adjustment problem. Value can be created by users or service demand, captured by a front-end company, retained by an issuer, burned by a protocol rule, accumulated by token holders, or absorbed by secondary-market buyers without ever becoming a payment to \W. Industry tokenomics work often describes this as value creation, capture, and accrual, and project documentation sometimes draws explicit money-flow or reserve-flow diagrams, such as GoodDollar's governance, reserve, and distribution architecture \cite{tokenomics2025,gooddollarDocs}. These value-flow views are useful, but Routed Closure asks a narrower and stricter question: does the external payment enter the reward loop through a verifiable route that pays the critical workers?

Put differently, value-capture analysis travels well to decentralized ecosystems only up to the point of value landing. In centralized firms and platforms, captured value usually remains inside a governable revenue pool that can be reallocated through contracts, budgets, subsidies, pricing, or managerial discretion. Decentralized ecosystems weaken this convergence condition. Captured value may land outside the reward loop, and the actor who captures value may not be the actor who performs the critical system function. Value-capture theory is therefore not wrong; it is under-specified once value landing and incentive payment are separated.

\subsection{Core Formalization}

The \emph{reservoir assumption} is the implicit assumption that captured value enters a centralized, governable pool from which the organization can fund continuing production, maintenance, subsidies, and rewards.

\textbf{Definition 1 (Pooled Capture).} Pooled Capture is value capture under the reservoir assumption: externally generated value is appropriated into a reallocable organizational pool controlled by a firm, platform, sponsor, or other focal allocator.

\textbf{Definition 2 (Route-Admissible Value).} For a critical incentive recipient \W{} over period \(T\), Route-Admissible Value is the captured external-use value that passes the route-admissibility gate for \W:
\[
\RAV_{\W}(T)
= \sum_f x_f(T)\,E(r_f)\,\mathbf{1}\{r_f \rightarrow \W\},
\]
where \(x_f(T)\) is an external-use value flow, \(E(r_f)\) is route enforceability, and \(\mathbf{1}\{r_f \rightarrow \W\}\) is one only when the route reaches the specified \W.

\textbf{Definition 3 (Routed Closure).} Routed Closure is the condition in which captured or generated external-use value is route-admissible for \W{} and is assessed against the relevant reward denominator:
\[
\RCR_{\W}(T)=\frac{\RAV_{\W}(T)}{V_{\W}(T)}.
\]
The first stage decides whether captured value counts; the second stage decides whether it is enough.

A capture-closure gap occurs when value is generated, appropriated, or made visible inside a decentralized ecosystem, but the value is not route-admissible for the relevant \W{} or remains insufficient relative to \(V_{\W}(T)\). The phrase does not deny value capture. It treats value capture as an intermediate observation that must pass route admissibility before it can support a coverage claim.

\begin{table}[htbp]
\centering
\caption{Pooled capture versus routed closure}
\label{tab:capture-closure-gap}
\footnotesize
\begin{tabularx}{\textwidth}{>{\raggedright\arraybackslash}p{3.3cm}YY}
\toprule
Dimension & Pooled capture / value-capture analysis & Routed Closure / \EVRC{} analysis \\
\midrule
Core question & Where is value appropriated, monetized, or accrued? & Can captured value count for \W, and is it enough? \\
Typical endpoint & Firm, platform, issuer, treasury, token holder, burn mechanism, or market value. & \RAV{} for a specified \W{} and a specified reward denominator. \\
Stage 1 & Value landing or appropriation is identified. & Route admissibility: landing node, route, enforceability, recipient, and period are accepted. \\
Stage 2 & Captured value can help estimate available resources. & Coverage adequacy: \(\RAV_{\W}(T)\) is compared with \(V_{\W}(T)\). \\
Main false positive & Value exists or is captured somewhere. & Captured value is counted as reward funding without route-admissibility evidence. \\
\bottomrule
\end{tabularx}
\end{table}

The key distinction is between \emph{discretionary routing} and \emph{protocol-enforced routing}. In a discretionary route, a firm or foundation can change take rates, subsidize contributors, redirect treasury funds, or stop doing so as strategy changes. In a protocol-enforced route, issuance, fees, transfers, escrow, staking rewards, or governance-executed payments are embedded in code, contracts, or formally executed rules. Protocol-enforced routes are more auditable, but also harder to repair when value lands in the wrong place. In this sense, blockchain value-flow design resembles a constitutional rule more than an ordinary operating policy: it fixes who receives value, under what conditions, and with what amendment cost. \EVRC{} therefore treats value capture as insufficient until value landing, route enforceability, beneficiary specificity, revocability, auditability, and reward coverage are separately coded.

This yields the paper's general proposition:

\begin{quote}
\emph{Value capture estimates potential value; route admissibility decides whether it counts for \W; coverage adequacy decides whether it is enough.}
\end{quote}

\subsection{Four Breakpoints}

\begin{description}
  \item[B1: Pseudo-consumption.] Payments motivated primarily by appreciation, yield, airdrops, or subsidy eligibility cannot be treated as ordinary external consumption demand.
  \item[B2: App--protocol fracture.] Value lands at an application, front-end, or company, but no verifiable route sends it to protocol-level rewards.
  \item[B3: Burn/capture mismatch.] Fees are burned or captured as scarcity, but \W{} is not paid by those fees.
  \item[B4: Issuance or market dependence.] Critical rewards rely mainly on inflation, subsidies, token price, or new buyers rather than external service payments.
\end{description}

\begin{table}[htbp]
\centering
\caption{Diagnostic types used in the contrast set}
\label{tab:diagnostic-types}
\footnotesize
\begin{tabularx}{\textwidth}{>{\raggedright\arraybackslash}p{3.1cm}YY}
\toprule
Diagnostic type & Included cases & Why it matters for \EVRC{} \\
\midrule
Unified revenue-pool baseline & YouTube & Shows a case where app-level revenue and creator payment can be mediated by one platform rule system. \\
App--protocol fracture & Steem/Steemit & Separates front-end/company finance from protocol issuance rewards unless a binding cash-to-\W{} route is shown. \\
Strong route, unresolved coverage & Bitcoin; Ethereum L1 & Fees or tips can have protocol routes while same-window coverage, burn treatment, MEV, issuance, and penalties remain separate tests. \\
Service or resource-payment candidate & Aave; Filecoin & User/service payment mechanisms can be positive route evidence, but still require denominators, risk treatment, and reconciliation before final \RCR. \\
Boundary or mismatch case & USDC; XRP & Issuer closure and token burn should not be recoded as host-chain or validator reward coverage. \\
\bottomrule
\end{tabularx}
\end{table}

\section{The EVRC Coding Protocol}

\EVRC{} operationalizes Routed Closure through a fixed coding order. The route-admissibility gate always precedes the coverage-adequacy gate:

\begin{enumerate}
  \item define the analysis unit: protocol, app, company, issuer, chain, DAO, or composite boundary;
  \item define the critical function and \W;
  \item classify payment motive: use-oriented, mixed financial service, investment-oriented, subsidy loop, or unknown;
  \item identify value landing: \(R_{\mathrm{app}}\), \(R_{\mathrm{proto}}\), \(R_{\mathrm{burn}}\), \(I_{\mathrm{new}}\), or another specified locus;
  \item code route type and routing strength \(E\);
  \item decide route admissibility: accepted into \(\RAV_{\W}(T)\), rejected, or source-blocked;
  \item determine the reward denominator \(V_{\W}(T)\) and whether coverage adequacy can be measured or bounded;
  \item assign evidence grade and claim gates.
\end{enumerate}

Routing strength \(E\) is ordinal: \(E=0\) means no identified route; \(E=0.25\) voluntary or discretionary route; \(E=0.5\) governance-mediated route; \(E=0.75\) strong off-chain contractual or platform-rule route; \(E=1\) protocol-enforced route. These values are not cardinal welfare scores. They are a discipline for preventing overclaims.

The band assignment uses four checks. \emph{Enforceability} asks whether the route can be compelled by code, law, escrow, or binding platform rules. \emph{Beneficiary specificity} asks whether the route names \W{} or a reward pool for \W, rather than a generic treasury. \emph{Revocability} asks whether the payer, team, or governance majority can stop the route without violating a binding rule. \emph{Auditability} asks whether an outside coder can observe the route, period, amount, and recipient. Governance-mediated payments are capped at \(E=0.5\) unless the vote has already created an escrowed, contractual, or protocol-executed payment rule; a governance attack or unilateral parameter change lowers the claim gate through revocability and risk treatment, rather than being hidden inside a higher route score.

Evidence grades are similarly conservative. Code, protocol rules, on-chain execution, audited filings, or source-captured official materials can support core mechanism claims. Official documentation and reliable dashboards can support mechanism or bounded numerical claims when dates and fields are specified. Media reports and narrative consensus cannot carry final closure claims.

The protocol also requires safe and forbidden claims. If the analysis unit is mixed, no final score is assigned. If \W{} is unclear, no \RCR{} is computed. If value capture fails route admissibility, it cannot enter \(\RAV_{\W}(T)\), even when the captured amount is real and well sourced. If payment motive is unclear, the claim is narrowed. If burn and \W{} payment are confused, the case is coded as B3. If issuance or market support dominates, the case is coded as B4.

B1 is coded before any revenue numerator is accepted. EVRC distinguishes use-oriented payments \(U\), financial-service use \(F\), mixed payments \(M\), investment-dependent payments \(I\), subsidy loops \(S\), and unknown motive \(X\). A conservative working numerator for external-use value is:
\[
V_{\mathrm{ext}}^{\mathrm{net}}
= U + F + \alpha M - \mathrm{rebates} - \mathrm{emissions} - \mathrm{wash/self\text{-}dealing},
\]
where \(0 \leq \alpha \leq 1\) is a disclosed haircut for mixed-motive activity. The formula is not a universal accounting identity; it is a coding guardrail. Gross volume, airdrop farming, subsidized wash activity, or self-dealing loops cannot enter the numerator merely because they create fees.

\section{Case Design and Current Evidence Boundaries}

The sample is not selected to produce attractive conclusions. It is selected to expose different forms of the capture-closure gap: centralized pooling, app--protocol fracture, security-budget routing, service-payment routes, resource-market extraction, issuer closure, and burn mismatch. Table~\ref{tab:cases} summarizes the current case boundaries.

\begin{table}[htbp]
\centering
\caption{Illustrative case roles and current claim boundaries}
\label{tab:cases}
\footnotesize
\begin{tabularx}{\textwidth}{>{\raggedright\arraybackslash}p{2.35cm}>{\raggedright\arraybackslash}p{2.75cm}YY}
\toprule
Case & Role & Current safe claim & Current forbidden claim \\
\midrule
YouTube & centralized baseline & Platform revenue can be pooled and redistributed to creators through company/platform rules. & Centralization is necessarily fairer or more sustainable. \\
Steem/Steemit & B2+B4 motivating case & Current captured sources identify no accepted cash-to-\W{} route from front-end/company revenue to authors, curators, witnesses, or the reward pool. & No revenue; universal historical route-null; final STEEM \RCR{}/\NCD. \\
Bitcoin & strong-route / weak-coverage contrast & Fees are protocol-routed to miners; recent baselines are low; the 2024 halving stress window shows extreme fee-share spikes. & Long-run security failure; final BTC \RCR{}/\NCD; stable fee replacement from one stress window. \\
Ethereum L1 & burn/validator split & Base-fee burn, tips, MEV, issuance, and penalties must be separated. Public routes support only partial subroutines so far. & Validator reward coverage from mixed burn/tip/issuance rows. \\
Aave & service-payment positive contrast & Borrower/user fees can route to suppliers and protocol revenue; aggregate rows support positive mechanism evidence. & Final AAVE token-holder closure or tail-risk coverage. \\
Filecoin & resource-market contrast & Storage market mechanism exists; provider deal revenue and reward fields are plausible extraction routes. & Accepted provider \RCR{} without active paid-deal numerator, reward denominator, and reconciliation. \\
USDC & issuer boundary & Reserve/redemption closure is issuer-level closure. & Host-chain validator reward coverage from issuer reserves. \\
XRP & burn mismatch & Transaction cost is burned, not paid to validators. & Burn as validator reward coverage. \\
\bottomrule
\end{tabularx}
\end{table}

Table~\ref{tab:coding-record} gives the minimum coding record used for the main-text cases. It is deliberately compact: the table records the field decisions needed to audit the claims, while the source register and case gates below determine whether any row can be upgraded to final \RCR{}/\NCD{} evidence.

\begin{table}[htbp]
\centering
\caption{Minimum EVRC coding record for the current main-text cases}
\label{tab:coding-record}
\scriptsize
\begin{tabularx}{\textwidth}{>{\raggedright\arraybackslash}p{1.55cm}>{\raggedright\arraybackslash}p{1.7cm}>{\raggedright\arraybackslash}p{1.8cm}>{\raggedright\arraybackslash}p{1.85cm}>{\raggedright\arraybackslash}p{1.75cm}YY}
\toprule
Case & Unit & \W{} & Landing & Route type & Current gate & Safe status \\
\midrule
YouTube & platform & creators & revenue pool & platform rule & baseline only & centralized comparator, not fairness claim \\
Steem & app/protocol & authors, curators, witnesses & company finance; issuance & no accepted cash-to-\W{} route in captured sources & final historical route-null blocked & bounded routing-boundary case \\
Bitcoin & chain & miners & protocol fees and subsidy & protocol-enforced fee route & stable fee replacement unresolved & strong route, low baseline, episodic spike evidence \\
Ethereum L1 & chain & validators & burn, tips, MEV, issuance, penalties & mixed protocol/proposer routes & fixed-window denominator blocked & mechanism-split case \\
Aave & protocol & suppliers, risk layers & user fees; protocol revenue & smart-contract service routes & final reward/risk coverage blocked & aggregate positive mechanism contrast \\
Filecoin & provider market & storage providers & deals, rewards, penalties & resource-market candidate routes & paid-deal/reward reconciliation blocked & mechanism anchored, extraction incomplete \\
USDC & issuer boundary & issuer operators; host validators excluded & reserves & issuer-level closure & host-chain reward route absent & issuer boundary case \\
XRP & ledger & validators & burned transaction cost & burn, not payment to \W{} & support/cost evidence absent & burn-mismatch case \\
\bottomrule
\end{tabularx}
\end{table}

\section{Steem/Steemit: The App--Protocol Fracture}

Steem is the paper's motivating case because it makes the prior payer question visible. Steem's reward fund and witness schedule are protocol-observable. Authors and curators compete for a post reward pool funded by blockchain issuance rules. Prior work documents decentralization limits, reward-pool abuse, decentralized curation security, and user incentives \cite{kim2019,li2019,kiayias2019,liu2022,li2021}. These findings describe how the internally funded pool is allocated and gamed. They do not, by themselves, answer who outside the token market pays for the pool.

Company-finance and chain-content sources show STEEM sales, cost management, and advertising-related activity at the Steemit/front-end layer. This is evidence of \(R_{\mathrm{app}}\) activity. Without a binding route, it is not evidence that advertisers or subscribers funded the protocol reward loop that paid creators, curators, or witnesses.

In the current audited source set, no accepted record simultaneously identifies external cash origin, binding route, \W{} recipient, and traceable period. The current source-acquisition checkpoint therefore records four statuses: accepted no-route-in-captured-sources evidence; no accepted final historical route-null proof; no accepted no-revenue claim; and no final STEEM \RCR{}/\NCD{} eligibility. The correct claim is bounded: no accepted cash-to-\W{} route has been identified in captured sources. The stronger historical claim--that no such route ever existed--remains source-coverage blocked.

\begin{figure}[htbp]
\centering
\begin{tikzpicture}[
  every node/.style={font=\footnotesize},
  box/.style={draw, rounded corners=2pt, align=center, text width=2.62cm, minimum height=0.78cm, fill=blue!4},
  reward/.style={draw, rounded corners=2pt, align=center, text width=2.62cm, minimum height=0.78cm, fill=green!6},
  warn/.style={draw, rounded corners=2pt, align=center, text width=2.62cm, minimum height=0.74cm, fill=red!5},
  arr/.style={-{Latex[length=2.2mm]}, thick},
  darr/.style={-{Latex[length=2.2mm]}, thick, dashed, red!70!black}
]
\node[box] (attention) at (0,0) {content and attention\\user activity};
\node[box] (cash) at (3.55,0) {front-end/company finance\\ads, STEEM sales, cost cuts};
\node[warn] (missing) at (7.25,0) {no accepted\\cash-to-\W{} route};
\node[reward] (issuance) at (3.55,-2.05) {protocol issuance rule\\reward fund};
\node[reward] (w) at (7.25,-2.05) {authors, curators, witnesses\\\W};
\node[box] (market) at (3.55,-3.80) {token-market absorption\\price, liquidity, expectations};

\draw[arr] (attention) -- (cash);
\draw[darr] (cash) -- (missing);
\draw[darr] (missing) -- (w);
\draw[arr] (issuance) -- (w);
\draw[arr] (market) -- (issuance);
\draw[darr] (cash.south) -- (issuance.north);
\end{tikzpicture}
\caption{The Steem/Steemit fracture. App or company finance is a value landing point, but current captured sources do not establish a binding route from front-end/company cash to the protocol reward loop.}
\label{fig:steem}
\end{figure}
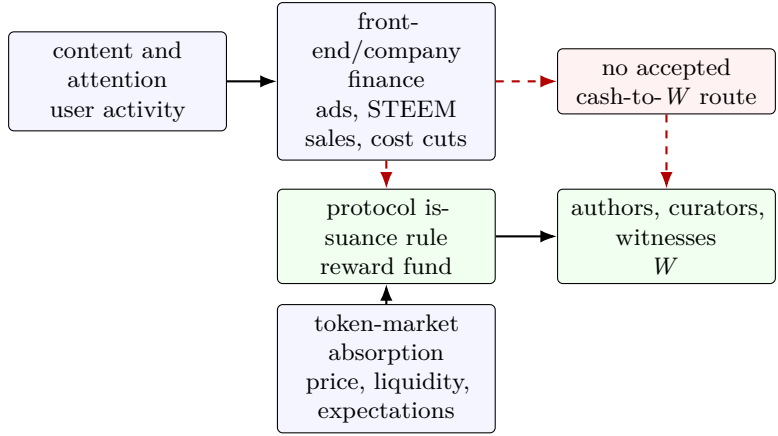

This is why Steem should not be described as a ``no revenue'' case. It is a routing-boundary case. The app/company layer can have activity while the protocol reward loop remains issuance- and market-dependent.

\section{Bitcoin, Ethereum, Aave, Filecoin, USDC, and XRP}

\subsection{Bitcoin and Ethereum}

Bitcoin has the cleanest routing type in the sample: transaction fees are protocol-routed to miners. This makes Bitcoin a direct instance of the security-budget literature's flow-payment problem \cite{budish2018,easley2019,huberman2021}. The unresolved question is coverage, persistence, and frequency. Current case records combine low recent baseline samples with a bounded high-fee stress window around the 2024 halving, where the reconstructed 144-block window reaches about 74\% fee share. The point for this framework paper is not to settle Bitcoin's long-run security budget. It is to show that strong route evidence and stable reward-coverage evidence are different tests.

Ethereum is the mechanism-split case. EIP-1559 makes base-fee burn analytically distinct from validator tips \cite{roughgarden2021}. For any fixed window \(T\), the validator-side reward object must be separated as:
\[
\begin{aligned}
R^{\mathrm{ETH}}_{W}(T)
= {} & \mathrm{priority\ fees}_{\rightarrow proposer}(T)
+ \mathrm{proposer\ MEV}(T) \\
& + \mathrm{consensus\ issuance}(T)
- \mathrm{penalties/slashing}(T).
\end{aligned}
\]
Base-fee burn belongs in a separate \(R_{\mathrm{burn}}(T)\) field. It may affect ETH scarcity or token-holder value, but it is not a payment to validator \W. A final validator \RCR{} would also need a same-window denominator and a recipient-consistent treatment of proposer-builder payments. Current public routes support partial execution-fee and head-reward subroutines, but not a final fixed-window validator \RCR{}.

\subsection{Aave and Filecoin}

Aave is a stronger service-payment mechanism contrast. Borrower/user fees can route to suppliers and protocol revenue. Current aggregate rows support a positive mechanism claim, but not final AAVE token-holder closure, incentives coverage, bad-debt coverage, or tail-risk sufficiency. In EVRC terms, Aave currently shows an accepted route mechanism and bounded aggregate numeric evidence, not final reward-coverage closure. This is not final AAVE token-holder closure.

Filecoin shows why resource-payment mechanisms still need extraction discipline. Storage-market documentation distinguishes storage clients, providers, storage fees, and block rewards. Candidate provider-level deal-revenue and reward fields could support \RCR{}, but current public/fallback routes do not yet provide accepted active paid-deal numerators, provider reward denominators, collateral/penalty treatment, and reconciliation. It is therefore a source-path and mechanism candidate, not a completed provider-coverage case.

\subsection{USDC and XRP}

USDC is an issuer-closure boundary. Reserve backing and redemption mechanisms are meaningful for issuer solvency, but reserve income does not automatically pay host-chain validators.

XRP is a burn-mismatch boundary. XRPL transaction costs are burned rather than paid to validators. Burn may affect supply, scarcity, or spam resistance, but it is not \W{} reward coverage.

\section{Relation to Prior Work}

The paper does not claim that prior literatures ignore sustainability, token design, business models, security budgets, DeFi revenue, or stablecoins. Rather, they address adjacent units of analysis.

Steemit studies are the closest threat. Prior work analyzes sustainable growth and token-economy design, reward allocation, decentralized curation, behavior, and operation extraction \cite{kim2019,li2019,kiayias2019,liu2022,li2021}. Routed Closure asks an upstream funding-route question: even if an internal reward system is well designed, does external use value reach the reward loop?

Token valuation models explain adoption, platform use, and token prices \cite{cong2021}. \EVRC{} separates token-price support from reward closure. If \W{} is paid mainly by newly issued tokens, dilution, or market absorption, token valuation may support nominal rewards without proving external payment to \W.

Business-model and ecosystem theory explain value creation and capture \cite{amit2001,teece2010,jacobides2018,weking2020}. The industry ``fat protocols'' thesis also made application-versus-protocol value distribution salient \cite{monegro2016}. Tokenomics design writing and project documentation increasingly discuss value creation, capture, accrual, reserve flows, and distribution flows \cite{tokenomics2025,gooddollarDocs}. Routed Closure does not replace those views; it specifies their missing post-capture test. Value-flow diagrams are not enough unless the route to the critical reward layer is specified and evidence-gated. Two projects can share a business model or tokenomics pattern while value lands in different places: a company balance sheet, protocol treasury, burn mechanism, liquidity supplier, validator, reward pool, or token market. The resulting gap is not ``no value capture.'' It is value capture without demonstrated reward closure.

Security-budget work is the mature special case. Prior work shows why fees, subsidies, congestion, and security expenditure must be separated \cite{budish2018,easley2019,huberman2021}. Routed Closure generalizes this payment-to-\W{} logic beyond miners and validators.

DeFi and stablecoin work shows that crypto systems can have real demand and cash-flow-like mechanisms \cite{schar2021,catalini2022,gorton2023}. These literatures also show why boundaries matter: Aave-style service fees do not automatically close token-holder tail risks, and USDC reserve income is issuer-level closure rather than host-chain validator closure.

\section{Limitations and Falsification}

The paper is a framework and protocol contribution, not a final empirical ranking.

\begin{enumerate}
  \item Steem has bounded current-source negative evidence, not final historical route-null proof. If a documented binding front-end cash route into protocol rewards is found, the Steem case must be recoded.
  \item Bitcoin has a complete halving stress-window candidate and long-run fee-share trend evidence, but no final BTC \RCR{}/\NCD{} or long-run security-budget verdict. Stable fee replacement requires non-overlapping current-regime windows, halving-regime comparison, and long-run route-to-fee evidence.
  \item Ethereum needs fixed-window priority-fee, MEV, issuance, penalty, burn, and denominator reconciliation before validator \RCR{} can be reported.
  \item Filecoin requires accepted active paid storage-deal payment numerators and provider reward denominators before provider \RCR{} can be reported.
  \item Aave currently supports mechanism and aggregate positive contrast evidence, not final token-holder or tail-risk closure.
  \item USDC and XRP are boundary cases, not final reward-coverage calculations.
  \item A true independent-coder pass has not yet been completed; therefore the current paper does not report inter-coder reliability.
\end{enumerate}

If \EVRC{} cannot consistently distinguish these cases under the same coding protocol, the contribution should be downgraded from a diagnostic framework to a conceptual essay.

\section*{Version, Scope, and Disclosure}

This arXiv v1 is a framework and protocol contribution with bounded comparative diagnostics. It is not a final empirical ranking: no final cross-case metric scores are reported, no final STEEM historical route-null claim is made, no long-run Bitcoin security-budget verdict is inferred, and the paper does not report inter-coder reliability. A fuller empirical crypto-economics extension is future work.

\noindent\textbf{AI/tool disclosure.} The author used AI-assisted tools to support literature organization, outline development, LaTeX editing, and language refinement. The conceptual claims, source selection, case interpretation, and final scholarly responsibility remain with the author.

\section{Conclusion}

The core claim of Routed Closure, operationalized through \EVRC, is simple: value capture is not reward closure. Decentralized sustainability analysis must trace whether external use value is route-admissible for the critical incentive layer and whether it covers the relevant reward denominator. Users, company revenue, token burns, asset prices, and protocol rewards are distinct objects. They should not be merged without route evidence.

Steem illustrates app--protocol fracture. Bitcoin illustrates strong routing with unresolved stable coverage. Ethereum illustrates mechanism splitting. Aave and Filecoin illustrate service-payment and resource-market contrasts. USDC and XRP illustrate issuer closure and burn mismatch. Across cases, the key question is not whether value exists somewhere in the ecosystem. It is whether value pays \W.

\section*{Data and Protocol Source Register}

Primary protocol and data sources are treated as mechanism or measurement evidence, not as scholarly prior work. The current working register separates eight source families: Steem whitepaper and developer/API materials for reward-fund and witness observability; mempool.space, Blockstream, and Coin Metrics routes for Bitcoin fee/subsidy rows and stress-window checks; EIP-1559 and ethereum.org documentation for Ethereum base-fee, priority-fee, issuance, reward, and penalty boundaries; Aave documentation and DefiLlama fee/revenue APIs for aggregate service-payment evidence; Filecoin, Lotus, and Spacescope materials for storage-market mechanism and provider-\RCR{} source gates; Circle transparency materials for USDC issuer boundaries; XRPL transaction-cost and validator-incentive documentation for the XRP burn boundary; and tokenomics/value-flow design sources used only to position the value-flow vocabulary. Source-gated rows remain bounded by their case gates: inclusion in this register does not make any final \RCR{}/\NCD{} ranking submission-ready.

\end{document}